\documentclass[aps,twocolumn]{revtex4-1}
\usepackage{bm}% bold math
\usepackage{amsmath}
\usepackage{hyperref}
\hypersetup{colorlinks=true, urlcolor=blue}

\usepackage{hyperref}
\usepackage{newlfont}
\usepackage{amssymb}
\usepackage{amsfonts}
\usepackage{amsmath}
\usepackage{graphicx}
\usepackage{bm}

\def \be{\begin{equation}}
\def \ee{ \end{equation} }

\begin{document}
\renewcommand*{\DefineNamedColor}[4]{%
  \textcolor[named]{#2}{\rule{7mm}{7mm}}\quad
  \texttt{#2}\strut\\}

\definecolor{red}{rgb}{1,0,0}
%\title{Stability against decoherence in macroscopic superpositions}
\title{Macroscopic Schr{\"o}dinger Cat Resistant to Particle Loss and Local Decoherence}

\author{Utkarsh Mishra, Aditi Sen(De), and Ujjwal Sen}

\affiliation{Harish-Chandra Research Institute, Chhatnag Road, Jhunsi, Allahabad 211 019, India}

\begin{abstract}
The Schr{\"o}dinger cat state plays  a crucial role in quantum theory, and has important fundamental as well as technological implications,
ranging from quantum measurement theory to quantum computers. 
The power of the potential implications of the cat state lies in the quantum coherence, as measured by the degree of entanglement,
 between its microscopic and  macroscopic sectors. 
We show that in contrast to other cat states, 
it is possible to choose the states of the macroscopic sector in a way that the resulting cat state, which we term as the W-cat state, has quantum coherence that is 
resistant to the twin effects of environmental noise -- local decoherence on all the particles and loss of a finite fraction of its particles. The states of the
 macroscopic sector of the W-cat state 
are macroscopically distinct in terms of their violation of Bell inequality.

\end{abstract}

\maketitle
%\section{Schrödinger cat}
%\section{Introduction and main results: The Schr{\"o}dinger cat}

%\textbf{checked portions: title, abs, para containing eq (1), secs II, III, IV.}

\section{Introduction and main results}

The recent developments in computation  and communication tasks have underlined the necessity to preserve quantum coherence in states shared by a 
large number of quantum systems \cite{rmp}.  
Feynman proposed that complex and large quantum systems can be efficiently simulated only by using a quantum computer \cite{Feynmann}. 
Shor's algorithm demonstrated that quantum algorithms can be used to efficiently solve problems that may not be possible with classical ones \cite{Shor}. 
To build a viable quantum computer that can compile and implement a quantum algorithm, which outperforms the ones running on classical machines, requires  quantum coherence preserved in a system of about \(10^3\) qubits \cite{Steane}. 
%
%One of main direction in answering such query is the discovery of algorithms by using quantum mechanics. Among several quantum algorithms,
% the first beakthrough came with the discovery of  Shor's algorithm \cite{Shor} -- factorizing integer in its prime factor with polynomial time. 
%It shows its power over its classical counterpart only when \(1000\) qubits can be prepared with coherence \cite{Steane}. 
Coherence in quantum states of a large number of particles 
is one of the essential ingredients for building a quantum communication network \cite{amaderreview}. 
Such exciting developments on the theoretical front were accompanied by several experimental proposals and realizations, by using e.g. photons, ion traps, cold atoms, and
nuclear magnetic resonance   \cite{experiments}.

Since preserving quantum coherence in states shared between multiparticle quantum systems is one of the basic necessities for many communicational and computational tasks, 
it is important to build a macroscopic entangled state. 
%Historically,  
Such an entangled state was first introduced by Schr{\"o}dinger in his seminal 1935 paper \cite{Sch01} through  the 
concept of the Schr{\"o}dinger cat, which is an entangled state between a microscopic system and a macroscopic one. 
The microscopic system can be an atom, that can decay spontaneously, with the undecayed state \(|\mbox{up}\rangle\)
and the decayed state \(|\mbox{down}\rangle\) making up a two-dimensional complex  Hilbert space (qubit). 
The macroscopic system was also conceived as a qubit made up of the alive and dead states of a cat, respectively denoted as 
\(|\mbox{alive}\rangle\) and \(|\mbox{dead}\rangle\). The quantum state of the combined micro-macro system is considered to be 
\begin{equation}
\frac{1}{\sqrt{2}} \left(|\mbox{up}\rangle |\mbox{alive}\rangle+|\mbox{down}\rangle|\mbox{dead}\rangle\right).
\end{equation}
Apart from its significance in technological pursuits, it is also important for understanding the quantum measurement problem and the quantum-to-classical transition \cite{Greenberger, Zurek01}.

The classic example 
%that is used for 
of the Schr{\"o}dinger cat is the 
  Greenberger-Horne-Zeilinger (GHZ) state \cite{GHZ}, given by
% shared between \(N+1\) particles,  
\begin{equation}
 |\mbox{GHZ}\rangle = \frac{1}{\sqrt{2}}\left(|0\rangle_{\mu} |0^{\otimes N}\rangle_{A_1 \ldots A_N} + |1\rangle_{\mu} |1^{\otimes N}\rangle_{A_1 \ldots A_N}\right).
\end{equation}
The party denoted by $\mu$ is the microscopic part of the Schr{\"o}dinger cat, while the parties \(A_1\) through \(A_N\) make up the macroscopic portion of the same.
The microscopic part is a single qubit, and is spanned by the orthonormal states \(|0\rangle\) and \(|1\rangle\). The macroscopic portion is built out of 
\(N\) qubits, denoted as \(A_1, A_2, \ldots, A_N\), with each being spanned by the orthonormal states \(|0\rangle\) and \(|1\rangle\).
%
%
%  was first introduced by Schr\"{o}dinger \cite{Sch01}, in his seminar paper in 1935, as a macroscopic state. In the GHZ state, the first part
%can be considered as a microscopic part or a cat which is alive  and  the second part  can be seen a macroscopic part or dead part. 
Noise effects on the GHZ state have been studied by using a variety of models \cite{Sim01}. However, as is well-known, the GHZ state loses \emph{all} quantum coherence if even a 
single qubit is lost.
% from its macroscopic sector. 
%
%
%Immense interests of macroscopic states in quantum information lead to a study of the decay rate of coherence of  such states under decoherence. 
%We address this question in the paper -- can we prepare a suitable cat state which will be robust against different noise models? Specifically, 
%we address the possibility of building different cat states which can preserve its coherence,  when they undergo loss of particles or local noises or both in the macroscopic part of 
%the system. We quantify the coherence in terms of entanglement as it is well established that the entanglement is the basic ingredients of different information theoretic protocols
%ranging from quantum communication \cite{rmp, amaderreview} to quantum computation \cite{HRoneway}. 
%It was shown that although the entanglement of a \(|GHZ\rangle\) state  can keep coherence upto certain percentage in some bipartitions under the local noise in the form of depolarizations 
%\cite{Sim01}, it  vanishes between microscopic and macroscopic systems with a single particle loss  or with local depolarizations. 

In this paper, we propose a macroscopic state, which we call the W-cat state, shared between \(N+1\) particles, given by 
\begin{eqnarray}
\label{HC-cat}
\arrowvert  H_{C}\rangle_{\mu A_{1}...A_{N}}  \quad \quad \quad \quad\quad \quad\quad \quad\quad \quad \quad \quad\quad \quad\quad \quad \nonumber \\
%& =& 
=
\frac{1}{\sqrt{2}}\left(\arrowvert 0 \rangle_{\mu} \arrowvert W_{N} \rangle _{A_{1}\ldots A_{N}}
%\nonumber\\&+& 
+
\arrowvert 1 \rangle _{\mu} \arrowvert 0 \ldots 0 \rangle_{A_{1} \ldots A_{N}}\right).
\end{eqnarray}
Here, \(|W_N\rangle\) is the \(N\)-particle W state \cite{wstate1, wstate2, wstate3, wstate4}, given by 
\begin{equation}
 \arrowvert W_{N} \rangle_{A_{1}\ldots A_{N}} = \frac{1}{\sqrt{N}} \sum \arrowvert 10 \ldots 0\rangle_{A_{1}...A_{N}},
\end{equation}
where the sum denotes an equal superposition of all \(N\) particle states consisting of a single \(|1\rangle\) and (\(N-1\)) \(|0\rangle\)s.
%Again, the party denoted by $\mu$ is the microscopic part of the Schr{\"o}dinger cat, while the parties \(A_1\) through \(A_N\) make up the macroscopic portion of the same.
%
%
%In this paper, we propose a macroscopic state, which we call W-cat state, between \(N+1\) particle, given by 
%\begin{equation}
% \label{HC-cat}
%|H_{c} \rangle = |0\rangle |W_N\rangle + |1\rangle |0\ldots 0\rangle, 
%\end{equation}
%where \(W_N = 1/\sqrt{N}(|10\ldots 0 + \ldots + |00\ldots 1\rangle)\).
 We show that the W-cat state is robust, i.e. can preserve quantum coherence in the form of entanglement between its micro and macro sectors,
 against  loss of a finite fraction of its particles and against local depolarizations on all its particles, and with the simultaneous action of both these noise effects. 
We then compare the robustness of this state with other macroscopic states, that can potentially be used as cat states. In particular, we find that 
for a finite number of particles in the macroscopic part, the W-cat state is more robust to local depolarizing noise than the GHZ state. 
%
%
%Apart from its fundamental importance,
Such investigations can potentially be a step towards building quantum memory devices using macroscopic systems.
%  in which microscopic state can be safely kept under the control of macroscopic particles. 

The paper is organized as follows.
In
Sec. \ref{sec-env}, we discuss the noise models that we have considered in this paper. 
The structure of the W-cat state that make it a Schr{\"o}dinger cat is revealed in Sec. \ref{sec-W-cat}. 
The entanglement measure used to measure the quantum coherence in the Schr{\"o}dinger cats is briefly discussed in Sec. \ref{sec-ent}. 
The effect of noise models (local decoherence and particle loss) on the W-cat states are discussed in Secs. \ref{sec-loss},  
\ref{sec-ld}, and  \ref{sec-loss+ld}. In 
Sec. \ref{sec-other}, we compare the W-cat state with other potential cat states. Finally, we present a conclusion  in
Sec. \ref{sec-conclu}.

%\section{Model of decoherence for general one  qbit}
\section{Environmental effects}
\label{sec-env}

It is of vital importance to investigate the effects of environmental noise on a Schr{\"o}dinger cat, in understanding its fundamental as well as technological implications.
Usually, the environmental effect that is considered for a Schr{\"o}dinger cat is decoherence. Here we consider the coherence properties of the 
Schr{\"o}dinger cat after it has been subjected, \emph{simultaneously} as well as separately, to local decoherence channels, in the form of 
local depolarizing channels, on all its constituent particles 
(in the micro as well as the macro sectors) and
 to loss of a finite fraction of its particles (in the macro part).

The depolarizing  channel destroys  off-diagonal  elements of  a  quantum density  matrix, destroying quantum coherence 
%resulting in decoherence 
in the
state, and is the usual model for decoherence phenomena \cite{depo-channel}. 
%This  phenomena  is  known  as decoherence. 
%This is also known as white noise because it is a random signal distributed over whole
%band width. In a system decoherence comes mainly because of its interaction to enviorment. 
Decoherence has, for example, been used to understand the 
%Hence the interaction and so the decoherence is 
%reason for 
transition of a quantum system to an effectively  classical system
%. Super selection or Einselection rule for wave function collapse
%are suppose to be consequences of such environment-induced decoherence
\cite{Zurek01}. Protecting a quantum system from decoherence is one of the  main 
challenges faced by quantum experimentalists and engineers.
% \cite{decoherence-in-qm}. 
%In fact decoherence is the main trouble in bringing quantum computer to the practice[give reference]. \\
Mathematically, the action of a  general depolarizing  channel, denoted by $D_{p}$, on a qubit is  given by \cite{depo-channel}
\begin{equation}
\label{dpc}
 \arrowvert i\rangle\langle j\arrowvert \to \frac{p}{2}I+(1-p) \arrowvert i\rangle\langle j\arrowvert,
\end{equation}
so that qubit remains unchanged with  probability $(1-p)$ and gets disturbed, by white noise, with probability $p$. 
Here, \(I\) denotes the identity matrix in the qubit Hilbert space. 
Note that this is a \emph{local} noise model, which is a natural choice 
for multiparty experimental situations, and our interest is in analyzing the coherence retained after the action of this local 
noise on all the qubits building up a Schr{\"o}dinger cat.
The cat state is also simultaneously inflicted by loss of some qubits. The effects of local decoherence and particle loss,  are also considered separately as special cases.

%and do not allow to choose preferred basis. Our interest is basically on such type of decoherence model and and we will analyze the effect
%of such decoherence on our cat state and compare the stability with other cat states. We will consider the more practical scenario where 
%the state can loose some particle during the preparation. We will then introduce local decoherence, as defined in $1$, to our state.

%\section{ New cat state, $C_{W}$ }
\section{The W-cat state}
\label{sec-W-cat}

In search for a cat state, that better withstands the environmental effects including particle loss, we propose the W-cat state, \(|H_C\rangle\).
The states \(|\mbox{alive}\rangle\) and \(|\mbox{dead}\rangle\) of the original Schr{\"o}dinger cat are certainly macroscopically different. 
In the ``GHZ-cat'' state, \(|GHZ\rangle\), these are replaced by the states \(|0^{\otimes N}\rangle\) and \(|1^{\otimes N}\rangle\) respectively. 
They are macroscopically different with respect to their magnetizations, i.e. with respect to their average values for the operator 
\(\frac{1}{N}\sum_{i=1}^N \sigma_z^i\), where \(\sigma_z^i= |0\rangle \langle 0| - |1\rangle \langle 1|\).
In case of the W-cat state, the states \(|\mbox{alive}\rangle\) and \(|\mbox{dead}\rangle\) of the original cat state are replaced respectively by the states 
\(|W_N\rangle\) and \(|0^{\otimes N}\rangle\). The latter are macroscopically different in terms of their violation of local realism. 

It is well-known that quantum mechanics is inconsistent with the assumption of an underlying hidden variable (``realistic'') theory that is also local. This is the 
statement of the celebrated Bell theorem \cite{Bell}. An important quantity quantifying the amount of violation of local realism by a quantum mechanical state 
is the 
critical visibility beyond which the state violates local realism. 
Among the ``alive'' and ``dead'' cat states that are used to build the W-cat state, the state \(|0^{\otimes N}\rangle\) certainly does not violate any 
local realism. However, the state \(|W_N\rangle\) has a critical visibility, given by \cite{wstate4}
\begin{equation}
p^{crit}_N = \frac{N}{(\sqrt{2} -1) 2^{N-1} + N}, 
\end{equation}
which tends to zero as \(N \to \infty\). It is in this sense that the states \(|W_N\rangle\) and \(|0^{\otimes N}\rangle\) of the 
macroscopic part of the W-cat state are macroscopically different.

%GHZ state turns out to be most popular choice. As we have already discuss the trouble with this state, it 
%become necessary to find out another kind of cat state(s), which can full fill more drastic situation. In this paper we have introduced a 
%new kind of state, which  we named $C_{w}$ state. This state is made of $N-$ party $W-$state, have maximal GGM and $N-$party zero qbit state 
%have zero GGM. So we can think of maximal GGM part as live cat and zero GGM part as death cat. The state has following form,   
%for which  t
The initial density  matrix, i.e. the density  matrix of the W-cat state before it passes through the local depolarizing 
 channels and is affected by particle loss,
 is denoted here by $\rho^{in}_{N+1}$, and
% The explicit  form  of  initial density  matrix  is 
 given by
\begin{eqnarray}
 \rho^{in}_{N+1}& = \frac{1}{2}\big[\arrowvert 0 \rangle  \arrowvert W_{N}\rangle\langle 0 \arrowvert \langle W_{N} \arrowvert
+ \arrowvert 0 \rangle \arrowvert W_{N}\rangle \langle 1\arrowvert \langle 0...0 \arrowvert\nonumber\\&
+ \arrowvert 1 \rangle \arrowvert 0...0 \rangle \langle 0\arrowvert \langle W_{N} \arrowvert
+\arrowvert 1 \rangle \arrowvert 0...0\rangle \langle 1 \arrowvert \langle 0...0\arrowvert\big].
\end{eqnarray}

%%%%%%%%%%%%%%%%%%%%%%%%%%%%%%%%%%%%%%%%%%%%%5

%%%%%%%%%Since we will be interested here in a situation where the system  can loose particles, we consider here the quantum state remaining, after we trace out  
%. To accomplish this fact in our analysis, we will trace out
%%%%%%%%% \(m\)
%%%%%%%%%%%% particles (equivalent to loss of \(m\) particles) from among the \(N\) particles constituting the macroscopic portion of the system.
% system $A_{1}...A_{N}$. 
%%%%%%%%%%The resultant  density matrix  will  then be an $(N-m+1)$ party system and 
%%%%%%%%%%%%5has the following  form:
%%%%%%%%%\begin{eqnarray}
%%%%%%%\rho^{in}_{N-m+1}&= \frac{1}{2}[\frac{(N-m)}{N} \arrowvert 0 \rangle \langle 0 \arrowvert \bigotimes  \arrowvert W_{N-m}\rangle \langle W_{N-m} \arrowvert\nonumber\\&+ \frac{m}{N}\arrowvert 0 \rangle\langle 0 \arrowvert\bigotimes\arrowvert 0...0\rangle  \langle 0...0\arrowvert \nonumber\\&+
%%%%%%%%%%%\sqrt{\frac{(N-m)}{N}}\arrowvert 0 \rangle \langle 1 \arrowvert \bigotimes \arrowvert W_{N-m}\rangle \langle 0...0\arrowvert\nonumber\\&+\sqrt{\frac{(N-m)}{N}}\arrowvert 1 \rangle \langle 0 \arrowvert \bigotimes \arrowvert 0...0\rangle \langle W_{N-m}\arrowvert\nonumber\\&+\arrowvert 1\rangle \langle 1 \arrowvert \bigotimes \arrowvert 0...0 \rangle \langle 0...0 \arrowvert].
%%%%%%%%%%%%\end{eqnarray}
%The resultant matrix is also a valid density matrix with trace unity, hermitian, and positive. Thus particle loose does not destroy the state 
%nature of quantum system. 

%%%%%%%%%%%%%%%%%%%%%%%%%%%%%%%%%%%%%%%%%%%%%%%%%%%%%%%%55

\section{Entanglement}
\label{sec-ent}
As mentioned before, it is important to understand the amount of environmental effect that a certain cat state can withstand. The ``cat-ness'' of a 
Schr{\"o}dinger cat is in the quantum coherence that exists between the micro and the macro sectors of the state. 
We measure the quantum coherence between these two sectors by using an entanglement measure. 

Entanglement of a two-party system shared between two parties \(A\) and \(B\), 
like the micro-macro system in the cat states, is defined as the inability of a quantum state of that system to be 
expressed in the separable form 
\begin{equation}
 \sum_ip_i\rho_A^i \otimes \rho_B^i,
\end{equation}
where \(\{p_i\}\) forms a probability distribution, and \(\rho_A^i\) (\(\rho_B^i\)) are states of the party \(A\) (\(B\)) \cite{Werner}. 
 
A convenient quantity to measure the entanglement in bipartite systems is the 
%Now  let's  study  the  entanglement of  this  state. We will use 
logarithmic negativity 
%as measure of entanglement.
%by  computing log  negativity 
\cite{vidal01}, defined, for a quantum state \(\rho\) of a two-party system, as 
%To compute log  negativity, we  have  to  find  out negativity. It turns  out the absolute value of sum of  negative eigenvalues of 
% partial  transpose of  density  matrix  with respect to  one  party. Let's  take  partial transpose with  respect  to $\mu$ system. Then 
%logarithmic negativity  is defined  as ,
\begin{equation}
 E_{N}(\rho) \equiv \log_2(2N(\rho)+1),
\end{equation}
where the ``negativity'', \(N(\rho)\), is given by the sum of the absolute values of the negative eigenvalues of the partial transpose of \(\rho\) with respect to 
either of the two parties. 
If either of the two parties is a qubit, the maximal value of \(E_N\) is unity.

\section{Entanglement of  W-cat  state with particle  loss}
\label{sec-loss}

In this section, we investigate the 
%We will interested here in a 
situation where the system looses a certain number of particles from its macroscopic part. 
%Since we will be interested here in a situation where the system  can loose particles, we consider here the quantum state remaining, after we trace out  
%. To accomplish this fact in our analysis, we will trace out
Suppose \(\rho^{in}_{N+1}\) looses \(m\)
%%%%%%%%%%%% particles (equivalent to loss of \(m\) 
particles from among the \(N\) particles constituting the macroscopic portion of the system.
% system $A_{1}...A_{N}$. 
%%%%%%%%%%
The resultant  density matrix  will  then be an $(N-m+1)$ party system
% and has 
having the following  form:
\begin{eqnarray}
\rho_{N-m+1}^{L}&= \frac{1}{2}\Big[\frac{(N-m)}{N} \arrowvert 0 \rangle \langle 0 \arrowvert \otimes  
\arrowvert W_{N-m}\rangle \langle W_{N-m} \arrowvert\nonumber\\&+ \frac{m}{N}
\arrowvert 0 \rangle\langle 0 \arrowvert\otimes\arrowvert 0...0\rangle  \langle 0\ldots0\arrowvert \nonumber\\&+
\sqrt{\frac{(N-m)}{N}}\arrowvert 0 \rangle \langle 1 \arrowvert \otimes \arrowvert W_{N-m}\rangle \langle 0\ldots 0\arrowvert\nonumber\\&
+\sqrt{\frac{(N-m)}{N}}\arrowvert 1 \rangle \langle 0 \arrowvert \otimes \arrowvert 0...0\rangle \langle W_{N-m}\arrowvert\nonumber\\&
+\arrowvert 1\rangle \langle 1 \arrowvert \otimes \arrowvert 0\ldots0 \rangle \langle 0\ldots0 \arrowvert\Big].
\end{eqnarray}
Here, the tensor product notation has been retained between the microscopic part (one qubit) and whatever has remained (\(N-m\) qubits) after the loss of \(m\) particles from the macroscopic part. 
%The resultant matrix is also a valid density matrix with trace unity, hermitian, and positive. Thus particle loose does not destroy the state 
%nature of quantum system. 
To investigate the effect of particle loss on the quantum coherence of the W-cat state $\arrowvert  H_{C}\rangle_{\mu A_{1}...A_{N}} $,
% under the effect of particle loss 
we find the entanglement of the resultant state (after particle loss)
in the 
 $\mu:A_1 \ldots A_{N-m}$ bipartition. Note here that we have assumed, without loss of generality, that the particles \(A_{N-m+1}\), \(A_{N-m+2}\), \(\ldots\), \(A_{N}\) are lost. 
After  taking  the  partial transposition with
respect to the microscopic sector of the system, the  partial transposed state of $\rho_{N-m+1}^{L} $ is seen to be  block-diagonal.
% The block,  which  gives us  negative 
%eigenvalue(s), will have $\frac{m}{N}$ as first entry. Rest  entries of first  row are $\sqrt{\frac{N-m}{N}}$ and same for  rest  
%entries of  first column. The  rest  entries  of  the matrix are  zeros. 
The  negative  eigenvalue of  the partial transposed state, denoted by $\lambda_{-}$, is  given by
\begin{equation}
 \lambda_{-} = -\frac{1}{2}\left(1 - \frac{m}{N}\right).
\end{equation}
Therefore, the entanglement of  W-cat 
%($| H_{C}\rangle $) 
state after the loss of \(m\) particles is given by
%logarithmic negativity, defined in  eqn$(6)$, for the  $C_{W}$ state is,  
\begin{equation}
\label{ekhon-7:32pm}
 E_{N}(\rho)=\log_2\left(2 -   \frac{m}{N}\right).
\end{equation}
%Where $N$ is the number of subsystems and $m$ the number of particle lost by the ,
 %where $m\ll N$. 
%The base of log is $2$.
 For large $N$, and for possibly large \(m\) satisfying $m\ll N$, the entanglement 
 between the microscopic and macroscopic subsystems reaches unity, irrespective of the value of \(m\), and 
hence the state is  (nearly) maximally entangled in this bipartition, as is also clear from Fig. $1$. The behavior of entanglement of the W-cat state with different 
  rates of particle loss and for different total numbers of particles,  is depicted in Fig. 1. 
 %depicted the  we have shown that how  entanglement varies as 
%the W state loses particles. 
\begin{figure}
\begin{center}
\includegraphics[height=0.2\textheight,width=0.35\textwidth]{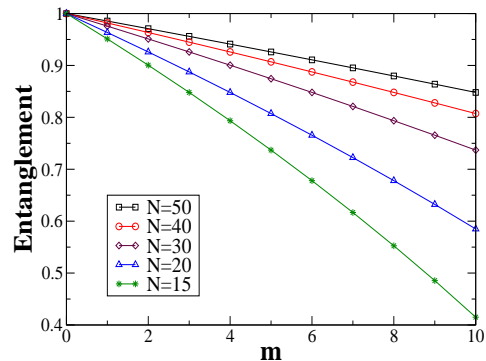}
%\scalebox{0.15}{\includegraphics{Nvsmforwstate}} 
\end{center}
\caption{(Color online.) Entanglement  after particle loss in W-cat state. 
%vs. number of particle loss by W-cat state. 
The horizontal axis represents  the number of particles lost  (\(m\))  from the macroscopic part of the W-cat state while 
the vertical one represents the entanglement between the micro and the macro parts of the W-cat state after particle loss. 
%quantified as logarithmic negativity. 
The entanglement with respect to \(m\) is plotted for different initial number of particles, \(N\). 
%From bottom to top, the total number of particles in W-cat state \(N\) are  respectively \(15\), \(20\), \(30\), \(40\) an d\(50\). 
The vertical axis is measured in ebits, while the horizontal one is in particles.
}
\end{figure}
The logarithmic decrease of entanglement with increasing numbers of particles lost, as seen from Eq. (\ref{ekhon-7:32pm}), is also clearly visible in Fig. 1.
%It is clear from the figure that the entanglement decreases logarithmically as one increases \( m\). 
%, . What is the relation of this decrease to entanglement of the
%different parts of $C_{W}$ the state?  

\section{Entanglement of W-cat state under local decoherence}
\label{sec-ld}

This section is devoted to the investigation of 
%e will analyze 
the entanglement properties of the W-cat state 
%$\arrowvert  H_{C}\rangle_{\mu A_{1}...A_{N}} $ state under 
under local decoherence effects on all the \(N+1\) constituent particles. To this end, each particle, whether from the microscopic or the macroscopic sector,
 of the initial state $\rho^{in}_{N+1}$ is fed to a 
%passes
%through 
the depolarizing channel, defined in Eq. (\ref{dpc}).  The output state, after this process, can be expressed as 
%form,
\begin{equation}
% \rho^{in}_{N+1}\mapsto 
D_{p}^{1}\otimes D_{p}^{2}\otimes \ldots \otimes D_{p}^{N+1}  \rho^{in}_{N+1} \equiv \rho^{DP}_{N+1},
\end{equation}
where $ D_{p}^{1}, D_{p}^{2},\ldots, D_{p}^{N+1}$ are $N+1$  depolarizing channels acting on the \(N+1\) particles in the initial state. 
%, each will act to the state, given in eqn($6$), according
%to the eqn($1$). Here too we will analyze entanglement in $1:2^{N-m}$
%cut. 
%The state in eqn($5$)can be, written afetr the channel and after the partial  transpose with respect to $1$ party, as
%\begin{eqnarray}
%\rho^{in}_{N-m+1}&= \frac{1}{2}[\frac{(N-m)}{N} D_{p}\arrowvert 0 \rangle \langle 0 \arrowvert D_{p} \bigotimes D_{p}^{{\bigotimes}N-m} \arrowvert W_{N-m}\rangle \langle W_{N-m} \arrowvert D_{p}^{{\bigotimes}N-m}\nonumber\\&+& \frac{m}{N}\arrowvert 0 \rangle\langle 0 \arrowvert \bigotimes D_{p}^{{\bigotimes}N-m}\arrowvert 0...0\rangle  \langle 0...0\arrowvert D_{p}^{{\bigotimes}N-m}\nonumber\\&+&
%\sqrt{\frac{(N-m)}{N}}\arrowvert 0 \rangle \langle 1 \arrowvert \bigotimes D_{p}^{{\bigotimes}N-m} \arrowvert W_{N-m}\rangle \langle 0...0\arrowvert D_{p}^{{\bigotimes}N-m}\nonumber\\&+&\sqrt{\frac{(N-m)}{N}}\arrowvert 1 \rangle \langle 0 \arrowvert \bigotimes D_{p}^{{\bigotimes}N-m} \arrowvert 0...0\rangle \langle W_{N-m}\arrowvert D_{p}^{{\bigotimes}N-m}\nonumber\\&+&\arrowvert 1\rangle \langle 1 \arrowvert \bigotimes D_{p}^{{\bigotimes}N-m}\arrowvert 0...0 \rangle \langle 0...0 \arrowvert D_{p}^{{\bigotimes}N-m}].
%\end{eqnarray}
The entanglement of the locally decohered W-cat state can now be analyzed in the micro : macro bipartition. The mathematical form of the entanglement will be presented in a more general context in the succeeding section, and so 
refrain from presenting it here. The results are depicted in Fig. 2, where we also present the corresponding curves for the GHZ states. Interestingly, we obtain that the W-cat state is more resistant to local 
decoherence than the GHZ state, and for example, for 10 particles in the macroscopic part, 
the W-cat state can preserve entanglement up to \(44\%\) of local decohering noise, while the GHZ state remain entangled until 
\(28\%\) of the same noise.

\begin{figure}
\begin{center}
%\scalebox{0.15}{\includegraphics{wvsGHZ}}
%\includegraphics[height=0.22\textheight,width=0.35\textwidth]{WvsGHZ} 
\includegraphics[height=0.25\textheight,width=0.4\textwidth]{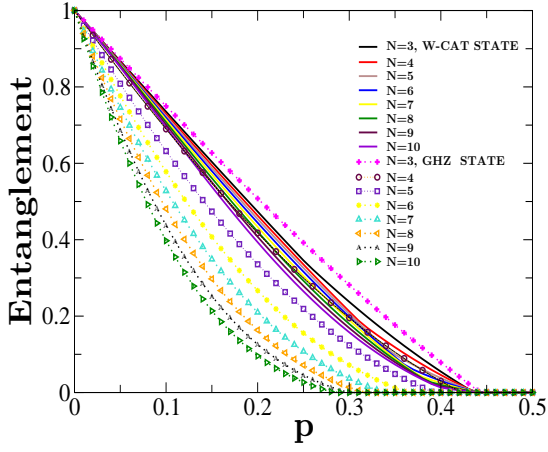} 
\end{center}
\caption{(Color online.) Entanglement of GHZ and  W-cat states against local decoherence. While the dotted lines are for the GHZ states, the 
continuous
 ones are for the W-cat. 
The horizontal axis represents the dimensionless (decohering noise) parameter \(p\), and the vertical axis is the entanglement in the micro : macro bipartition (in ebits). 
%Comparison of entanglement between W-cat state and GHZ state under local depolarizations.
}
\end{figure}

\begin{figure}[h]
\begin{center}
\includegraphics[width=0.4\textwidth]{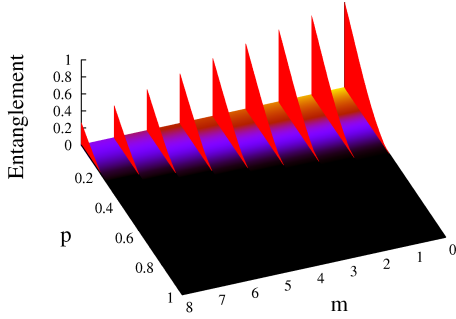}
\end{center}
\caption{(Color online.) Effect of local decoherence and particle loss on W-cat state.
Entanglement (measured in ebits) in the micro : macro bipartition is plotted on the vertical axis against 
a base of the dimensionless depolarizing parameter \(p\), and the number of lost particles (\(m\)). 
%The effect of entanglement of W-cat state under decoherence: Th entanglement is plotted against the loss of particles and the local depolarizing parameter 
%\(p\) for \(N=10\).
The W-cat state under consideration is of 11 qubits, so that \(N=10\).
}
\end{figure}

\begin{figure}[h]
\begin{center}
\includegraphics[width=0.4\textwidth]{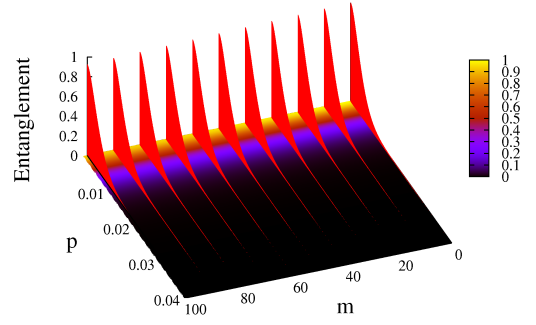}
\end{center}
\caption{(Color online.) Effect of local decoherence and up to 10\% particle loss on a W-cat state of \(10^3 + 1\) qubits. 
All other considerations except \(N\) is the same as in Fig. 3. 
%The behavior of entanglement with respect to the particle loss and local decoherence for macroscopic system with \(N=1000\).
}
\end{figure}

\section{Entanglement of W-cat state under particle loss and local decoherence}
\label{sec-loss+ld}
%This is the main section of this article.

We now consider the situation where the W-cat state is affected by local decoherence as well as by particle loss. We assume that \(m\) partcles are lost (from the macro part) and that the remaining 
\(N-m+1\) particles are all affected by local decoherence as modelled by the depolarizing channel. The entanglement in the micro : macro bipartition is analyzed for the resulting \(N-m+1\)-party state. The two eigenvalues 
of the partial transposed state that make the maximum contribution are 
$\lambda_{-}^{(1)}$ and $\lambda_{-}^{(2)}$, where $ |\lambda_{-}^{(1)}| > |\lambda_{-}^{(2)}|$.
These two eigenvalues are given by
\begin{eqnarray}
\lambda_{-}^{(1)}= \frac{1}{4}\Big\{c+(N-m-1)d+a \quad \quad \quad \quad\quad \quad\quad \quad\quad \quad\nonumber\\-\sqrt{4(N-m)b^{2}+(c+(N-m-1)d-a)^{2}}\Big\},\quad \quad
\end{eqnarray}
where
\begin{eqnarray}
a&=&\gamma_{1} \tilde{p} +\frac{m}{N}{\tilde{p}}^{N-m+1}+\frac{p}{2}{\tilde{p}}^{N-m},\nonumber\\
b&=&\frac{1}{\sqrt{N}}(1-p)^{2}{\tilde{p}}^{N-m-1},\nonumber\\
c&=&\alpha_{1}\frac{p}{2}+\frac{m}{N}\left(\frac{p}{2}\right)^{2}{\tilde{p}}^{N-m-1}+\frac{p}{2}{\tilde{p}}^{N-m},\nonumber\\
d&=&\frac{1}{N}\frac{p}{2}(1-p)^{2}{\tilde{p}}^{N-m-2},
\end{eqnarray}
with
$\alpha_{1} = \frac{1}{N}\left({\tilde{p}}^{N-m}+(N-m-1) (\frac{p}{2})^{2}{\tilde{p}}^{N-m-2}\right)$,
$\gamma_{1}=(\frac{N-m}{N})(\frac{p}{2}){\tilde{p}}^{N-m-1}$, \(\tilde{p}=1-\frac{p}{2}\),
and
\begin{eqnarray}
%\lambda_{-}^{(2)}&=&\frac{1}{4}(-\sqrt{4(\tilde{N}+2)e^{2}+(-a_{1}+b_{1}+f+\tilde{N}g)^{2}}\nonumber\\&+&(a_{1}-b_{1}+f+\tilde{N}g) !!!!\\
\lambda_{-}^{(2)}=\frac{1}{4}\Big\{(a_{1}-b_{1}+f+{\tilde{N}}g)\quad \quad \quad \quad\quad \quad \quad \quad\quad \quad\nonumber\\
-\sqrt{4({\tilde{N}}+2)e^{2}+(-a_{1}+b_{1}+f+{\tilde{N}}g)^{2}}\Big\}, \quad \quad
\end{eqnarray}
where
\begin{eqnarray}
a_{1}&=&\left(\frac{p}{2}\right)^{2}{\tilde{p}}^{N-m-1}+\frac{m}{N}\frac{p}{2}{\tilde{p}}^{N-m}+\gamma_{2}{\tilde{p}}, \nonumber \\
b_{1}&=&\frac{1}{N}(1-p)^{2}{\tilde{p}}^{N-m-1}, \nonumber \\
e&=&\frac{1}{\sqrt{N}}(1-p)^{2}\frac{p}{2}{\tilde{p}}^{N-m-2}, \nonumber \\
g&=&\frac{1}{N}(1-p)^{2}\left(\frac{p}{2}\right)^{2}{\tilde{p}}^{N-m-3}, \nonumber \\
f&=&\left(\frac{p}{2}\right)^{2}{\tilde{p}}^{N-m-1}+\frac{m}{N}\left(\frac{p}{2}\right)^{3} {\tilde{p}}^{N-m-2}+\alpha_{2}\frac{p}{2}, \nonumber \\
\end{eqnarray}
with
$\alpha_{2}= (1/N)[2{\tilde{p}}^{N-m-1}(p/2)+(N-m-2)(p/2)^{3}{\tilde{p}}^{N-m-3}]$,
%\textbf{!!!!!!!!!!!!!!!!!!!!!!!!!!!!!!!!!!!!!!!!!!!!!!!!!!!!!!!!!!!!!!!!!!!!!!!!!!!!!!!!!!!!!!}\frac{1}{N}(2\frac{p}{2}{\tilde{p}}^{N-m-1}+(N-m-2)\left(\frac{p}{2}\right)^{3}{\tilde{p}}^{N-m-3})$,
$\gamma_{2}=\frac{1}{N}({\tilde{p}}^{N-m}+(N-m-1)\left(\frac{p}{2}\right)^{2}{\tilde{p}}^{N-m-2})$, \(\tilde{N} = N-m-4\).
The remaining eigenvalues make a contribution to the logarithmic negativity that is rather insignificant, and so for \(N=8\), \(m=1\) and 
\(p=0.1\), their contribution to the entanglement is less than \(10^{-2}\). 
Note here that by setting \(m=0\), we can obtain the entanglement expressions for the case considered in the preceding section.

The entanglements are plotted in Figs. 3 and 4. In particular, in Fig. 4, we consider the case when the macroscopic system is constituted out of \(N=10^3\) particles, and we find that the entanglement in the micro : macro bipartition remains almost at 
its initial maximal value even with the loss of about \(10\%\) of its particles. 
Entanglement remains nonzero even when the remaining 90\% particles are fed to local depolarizing channels until \(p \lesssim .03\).

%We are looking for more worst situation where a cat state can sustain multipartite entanglement
%even after loss of few particle, say $m$, where $m\ll N$, the number of parties involved. For such cases, GHZ state remains no longer a 
%popular choice. We have checked this for $C_{w}$ state and found that even for $N=1000$, $C_{w}$ state remains multipartite entangled after 
%loosing few particles. As we can see from the graph that $C_{w}$ state don't care, if it looses few particles. For large $N$ it remain maximally
%multipartite entangled even after loosing few particles. This feature also make $C_{w}$ state more practical as compared to GHZ state.
%\textbf{We have to mention the highest negative eigenvalues in this section}

%\section{Entanglement of some other states}
\section{Noise effects on Entanglement of other cat states}
\label{sec-other}

In this section, we consider some other states which may potentially be considered as cat states, and compare their ability to withstand particle loss.
%Here we have considered some other states, and compare their stability with respect to $H_{C}$ state.
We begin by considering the state 
%First we will consider the following state defined as,
\begin{eqnarray}
 \arrowvert  \Psi_{1}\rangle _{\mu A_{1}...A_{N}} = \frac{1}{\sqrt{2}}\left(\arrowvert 0 \rangle _{\mu} \arrowvert W_{N} \rangle _{A_{1}...A_{N}}
%\nonumber\\&+& 
+
\arrowvert 1 \rangle _{\mu} \arrowvert \widetilde {W} \rangle_{A_{1}...A_{N}}\right), \nonumber \\
\end{eqnarray}
where 
%\begin{equation}
% \arrowvert W_{N} \rangle _{A_{1}...A_{N}} = \frac{1}{\sqrt{N}} \sum_{p}\arrowvert 10...0\rangle.
%\end{equation}
%and
\begin{equation}
 \arrowvert \widetilde W_{N} \rangle _{A_{1}...A_{N}} = {\sigma_x}^{^\otimes N} \arrowvert W_{N} \rangle _{A_{1}...A_{N}}, 
%\frac{1}{\sqrt{N}} \sum_{p}\arrowvert 01...1\rangle.
\end{equation}
where \(\sigma_x = |0\rangle \langle 1| + |1\rangle \langle 0|\).
%where sum is over all permutation.  
The state \(\arrowvert \ \Psi_{1}\rangle\) is the G state of Ref. \cite{Gdansk}.
This state is a cat state in the sense that the states \(|W\rangle\) and \(|\widetilde{W}\rangle\) are macroscopically distinct in terms of their 
\(\sigma_z\)-magnetizations, just like in the case of the GHZ state.
 This state, however, becomes separable if,
for any $N$, we lose more than two particles. 
Another state that can be considered as the cat state is apparently quite similar to the W-cat state, with only the 
N-qubit W state state replaced by the state  \( \arrowvert \widetilde W_{N} \rangle\).
%The other state which we have considered is the state similar to $H_{C}$, only $ W$ has been replaced by $\widetilde W$. 
%After this change,
%The state becomes,
This state is given, therefore,  by 
\begin{eqnarray}
 \arrowvert  \Psi_{2}\rangle _{\mu A_{1}...A_{N}} = \quad \quad \quad \quad \quad \quad \quad \quad \quad \quad \quad \quad \quad \quad \quad \quad \quad \quad \nonumber \\
\frac{1}{\sqrt{2}}\left(\arrowvert 0 \rangle _{\mu} \arrowvert \widetilde W_{N} \rangle _{A_{1}\ldots A_{N}} 
+ \arrowvert 1 \rangle _{\mu} \arrowvert 0 \ldots 0 \rangle_{A_{1}\ldots A_{N}}\right). \quad \quad 
\end{eqnarray}
%Where $\widetilde W$
This state is a cat state in the same sense as the W-cat state -- the Bell inequality violations of \( \arrowvert \widetilde W_{N} \rangle\) and 
\(\arrowvert 0 \ldots 0 \rangle\) are drastically different. Moreover, the states \( \arrowvert \widetilde W_{N} \rangle\) and 
\(\arrowvert 0 \ldots 0 \rangle\) are also macroscopically different in terms of their \(\sigma_z\)-magnetizations.
% that both of its parts are magnetize in difference direction. The first has all
%its qubits up, except one, and second part has all its qubits down. 
This state becomes separable if, for any $N$,  we lose more than one 
 particle.
An interesting generalization of the GHZ state is the concatenated GHZ state \cite{Dur01}, and is given by 
%. This has following form,
\begin{equation}
 \arrowvert\Psi_{3}\rangle = \frac{1}{\sqrt{2}}\left(\arrowvert GHZ_{l}^{+}\rangle^{\otimes (N+1)}+\arrowvert GHZ_{l}^{-}\rangle^{\otimes (N+1)}\right),
\end{equation}
where
\begin{equation}
 \arrowvert GHZ_{l}^{\pm}\rangle= \frac{1}{\sqrt{2}}\left(\arrowvert 0 \rangle ^{ \otimes l}\pm\arrowvert 1 \rangle ^{ \otimes l}\right).
\end{equation}
Here there are \(N+1\) logical qubits, and each logical qubit is built by using \(l\) physical qubits. 
Loss of all physical qubits from a logical qubit renders this state separable, like the GHZ state. Also, loss of some physical qubits from different logical qubits
leads to separable states. 
%also renders 
%Where, $N$  is the number of logical qubits, and each logical qubit is build by $l$ physical qubits. This state is locally equivalent
%to GHZ state with $Nl$ subsystem.  The state become separable if we lose one physical qubit from each logical qubits. 
%So we find that, while the state defined  in eqn($15$) is stable under local decoherence it can not sustain particle lose. The entanglement 
%of other two states defined in eqn($11$) and eqn($14$) vanishes under particle lose. 

\section{Conclusion}
\label{sec-conclu}
The concept of the Schr{\"o}dinger cat is an important aspect of quantum physics with significance on the fundamental front as well as in useful applications.
There is an ongoing effort towards experimental realization of such cat states in a variety of physical substrates.
% \cite{expt-cat}.
We have proposed a Schr{\"o}dinger cat, whose ``alive'' and ``dead'' states are modelled by two quantum states that drastically differ by their 
amounts of violation of Bell inequalities. We show that this state is robust against loss of a finite fraction of its particles and simultaneously against 
local depolarizing channels, modelling a local decoherence mechanism. We compare our results with other potential cat states, including the Greenberger-Horne-Zeilinger
state.

% large body of research

%So we have propose a cat state, $C_{w}$, which can sustain particle loose as well as decoherence.
%\textit{Acknowledgement}- 

%begin{figure}[h]
%\includegraphics[width=1.65 in]{combi2dN12_TC}
%\includegraphics[width=1.65 in]{combi2dN12_CC}
%\resizebox{3.25 in}{!}{
%\includegraphics{combi2dN12_Dis.eps}\hspace{1cm}
%\includegraphics{combi2dN12_LN.eps}}
%\caption{(Color online.) Ergodicity versus nonergodicity on 2D \(XY\) models. 
%QD (left, in bits) and LN (right, in ebits) for a 2D array of 12 spins with 
%the two-dimensional quantum transverse \(XY\) model of 12 spins with 
%PBC. 
%All considerations are the same as in Fig. \ref{fig:1dinfall}. 
%Depictions used here are same as in Fig. \ref{fig:1dinfall} and Fig. \ref{fig:1dn12}. Complementary statistical mechanical behavior between entanglement-based measures and information-theoretic measures are also observed in this case.
%Quantum correlations as a function of temperatures for two dimensional quantum $XY$ model with $N=12$ spins is plotted here. Quantum discord (($Q$, left)) and quantum entanglement ($E_N$, right) and  have contradictory statistical behavior for ergodicity. We have plotted the values of the corresponding physical quantity, for evolved sate, as horizontal lines for different values of the applied transverse fields viz., $a/J=0.2$ (red squares), $a/J=0.6$ (blue diamonds), and $a/J=2$ (pink triangles). \(a/J\) is dimensionless quantity.
%}
%\label{fig:2dn12}
%\end{figure}

%end{thebibliography}
\end{document}